\documentclass[aps,twocolumn,prl,color,psfig,epsf,nobalancelastpage]{revtex4-1}
\usepackage{amsmath}
\usepackage{color}
\usepackage{amsfonts}
\usepackage{epsf}
\usepackage{graphicx}
\usepackage{epstopdf}
\usepackage{ulem}
\usepackage{balance}

\begin{document}

\title{Electrostatic inactivation of RNA viruses at air-water and liquid-liquid interfaces}
\author{C.~A. Brackley, A. Lips, A. Morozov, W.~C.~K. Poon, D. Marenduzzo}
\affiliation{SUPA, School of Physics and Astronomy, The University of Edinburgh, Peter Guthrie Tait Road, Edinburgh, EH9 3FD, Scotland, United Kingdom}

\begin{abstract}
Understanding the interactions between viruses and surfaces or interfaces is important, as they provide the principles underpinning the cleaning and disinfection of contaminated surfaces. Yet, the physics of such interactions is currently poorly understood. For instance, there are longstanding experimental observations suggesting that the presence of air-water interfaces can generically inactivate and kill viruses, yet the mechanism underlying this phenomenon remains unknown. Here we use theory and simulations to show that electrostatics provides one such mechanism, and that this is very general. Thus, we predict that the free energy of an RNA virus should increase by several thousands of $k_BT$ as the virion breaches an air-water interface. We also show that the fate of a virus approaching a generic liquid-liquid interface depends strongly on the detailed balance between interfacial and electrostatic forces, which can be tuned, for instance, by choosing different media to contact a virus-laden respiratory droplet. We propose that these results can be used to design effective strategies for surface disinfection. Intriguingly, tunability requires electrostatic and interfacial forces to scale similarly with viral size, which naturally occurs when charges are arranged in a double-shell distribution as in RNA viruses like influenza and all coronaviruses.
\end{abstract}

\maketitle

The physics of virus-surfaces and virus-interface interactions is ripe with interesting experimental observations~\cite{Gerba1984,Castano2020}, but several of these lack a satisfactory theoretical understanding. For instance, the number of viable viruses on a surface typically decays over time as $n(t)=n_0 e^{-t/t_0}$, with the timescale $t_0$ ranging from hours to days~\cite{vanDoremalen2020}. A variety of factors affect $t_0$, such as the nature of the surface and the absolute humidity~\cite{Shaman2009}. However, a mechanistic understanding of this timescale is currently lacking. Equally intriguingly, several reports suggest that viruses are inactivated whenever they are exposed to flow in the presence of air/liquid/solid interfaces, for instance when air is bubbled through a viral solution \cite{Adams1948}, or when such a solution is tumbled in a test tube~\cite{Trouwborst1974} or passed through a packed bed of beads~\cite{Thompson1999}. Viral inactivation by exposure to suitable interfaces is clearly potentially relevant to surface disinfection; yet, once again, the physical mechanisms are unknown. 

A potentially relevant observation is that viruses are highly charged. This is both because they contain nucleic acids (RNA or DNA) with  high negative charge -- one electron at each phosphate -- and because the proteins constituting their capsid shell have a pH-dependent charge, which can be as high as one electron per nm$^2$~\cite{Siber2007,Siber2012,vanderSchoot2005}. {\it A priori}, then, we expect electrostatics to be important to viral energetics. Indeed, the electrostatic energy stored in viral capsids and genomes is estimated to be $\lesssim 10^4 k_BT$ for typical RNA viruses~\cite{Siber2012}. In a physiological buffer, this repulsive energy can be offset by van der Waals and hydrophobic attractions, thus enabling the virion to self assemble spontaneously. 

Near an air-water interface, the electrostatic Debye-H\"uckel repulsion between two point-like charges switches from an exponentially screened interaction to an unscreened and long-range effective dipolar repulsion~\cite{Pieranski1980,Hurd1985,Netz1999,Frydel2007}, whereas charges wholly in the air phase interact via the Coulomb potential. Thus, electrostatic repulsion generically increases close to an air-water interface, and may overcome attractive interactions to cause viral destabilisation. The free energy increase needed to disassemble a virus is likely relatively small, because disassembly is a necessary part of the infection cycle~\cite{Smyth2002}. We therefore hypothesize that strong electrostatic interactions at interfaces may lead to viral deactivation. 

To test this hypothesis, we solve numerically the non-linear Poisson-Boltzmann (PB) equation for a viral particle approaching an air-water or a liquid-liquid interface. This reveals a significant electrostatic free energy cost opposing adsorption to the interface. We compare this cost with the energy gained when a nanoparticle covers part of an interface, which is responsible for the stabilisation of Pickering emulsions~\cite{Pickering1907,Ramsden1903}. Depending on physical parameters such as the dielectric constants and Debye lengths in the two contacting media, we find that the competition between the electrostatic and Pickering effects yields a transition between a regime where the virion breaches the interface spontaneously, and one where it is repelled from it. These findings shed light on previous viral inactivation experiments~\cite{Adams1948,Trouwborst1974,Thompson1999}, and suggest strategies for effective surface decontamination. Our calculations are distinct from those in previous work, which focussed on solid surfaces rather than fluid interfaces~\cite{Gerba1984}.


\subsection*{A Poisson-Boltzmann model for an RNA virus close to an interface}

\begin{figure}[t]
\centerline{\includegraphics[width=0.5\textwidth, angle=0]{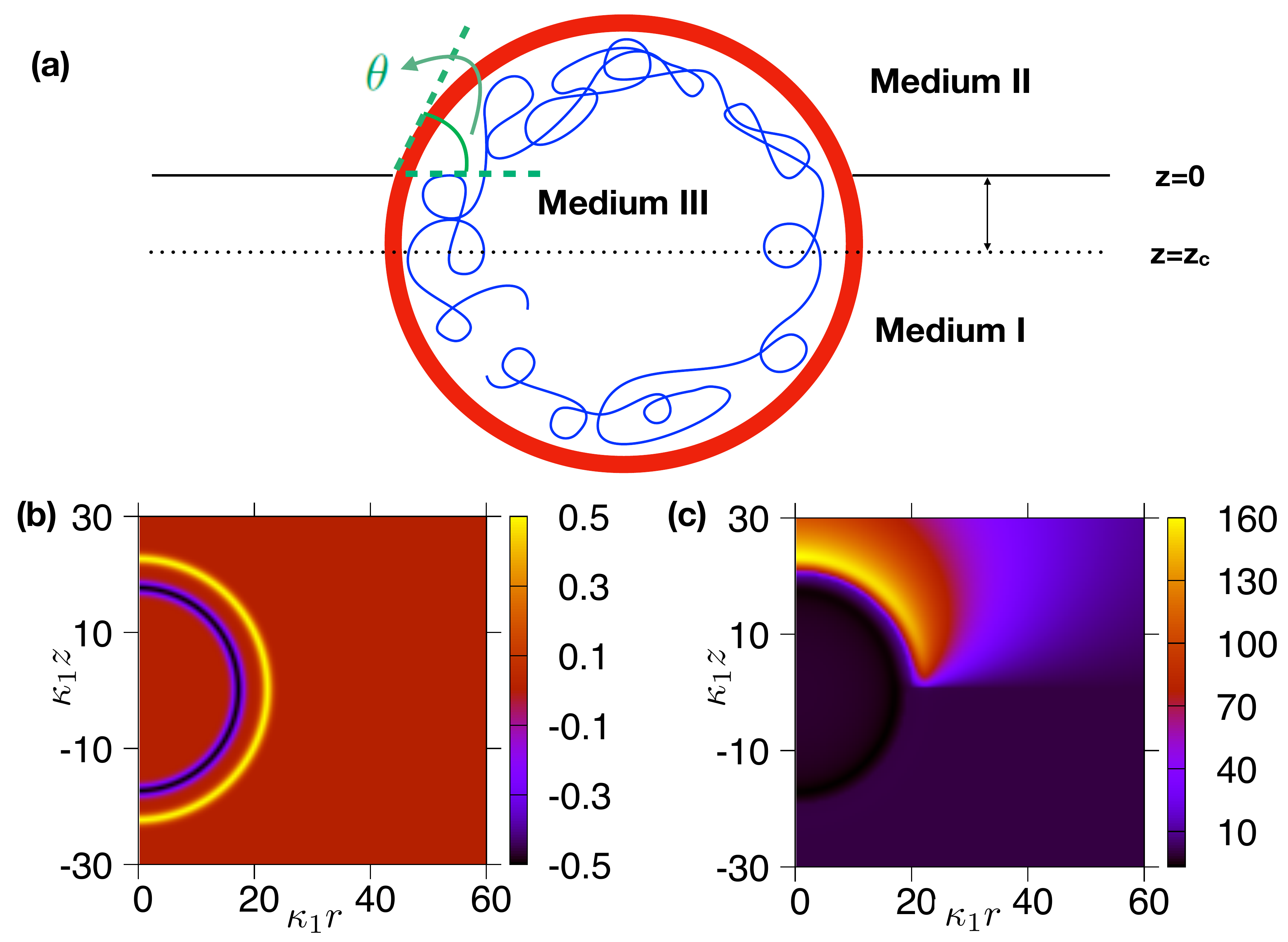}}
\caption{(a) Schematics of the system under consideration. The capsid shell and folded RNA of a virion are modelled as two concentric shells with charge density $-\sigma$ and $\sigma$ respectively. (b) Model charge distribution used in numerical simulations of a virion at an air-water interface ($\kappa_2/\kappa_1=0$, $\epsilon_2/\epsilon_1=1/80$, $z_c=0$). The heat map gives the dimensionless local charge density (see Appendix). (c) Corresponding potential field found by numerically integrating the non-linear PB equation, Eq.~(\ref{interfacePB}), for an air-water interface. The heat map gives the value of the dimensionless electrostatic potential $\tilde{\phi}$. }
\label{fig1}
\end{figure}

In an RNA virus, the flexible (negatively charged) RNA is adsorbed to the (positively charged) interior of the protein capsid. We model this by two oppositely-charged concentric spherical shells of average radius $R$ and with spacing $2\delta$ between them (Fig.~\ref{fig1}a). For simplicity, we consider an equal charge density, $\sigma$, for both shells, so that the viral particle carries a net charge (i.e., it has a non-zero charge monopole). Selected numerical simulations and theoretical arguments (see Appendix) suggest that the trends we find are unaffected if the charge densities are tuned to give a neutral virion. We consider a planar interface separating media I and  II with inverse Debye lengths $\kappa_1$ and $\kappa_2$ and permittivities $\epsilon_1$ and $\epsilon_2$. 

We introduce cylindrical spatial coordinates $z$ (height with respect to the interface plane) and $r$ (perpendicular distance to $z$ axis). The centre of mass of the viral particle lies at $r=0$ and $z=z_c<0$. In all our numerical calculations medium I ($z<0$) is an aqueous physiological buffer, which we model as a $150$ mM monovalent salt solution with $\kappa_1^{-1}\sim 1$ nm and $\epsilon_1\sim 80\epsilon_0$, where $\epsilon_0$ is the dielectric permittivity of vacuum. The interior of a capsid (medium III in Fig.~\ref{fig1}) is likely a different electrostatic environment from the aqueous surrounding. We set $\kappa_3=0.1 \kappa_1$ (as the capsid may only be partially permeable to salt) and $\epsilon_3=\epsilon_1/16$~\cite{Siber2012}. 

Our concentric shell model of a virion is highly simplified. The real protein capsid charge is pH-dependent. Moreover, the exterior walls of the capsid tend to be oppositely charged, but modelling the capsid as a uniformly charged shell gives similar results in PB modelling of the bulk electrostatics~\cite{Siber2012}. We also note that we model virions of a fixed shape, which is a good approximation until they are subjected to forces of $\sim 1$ nN, at which point capsids deform or rupture~\cite{Michel2006}. 


With monovalent salts in both medium I and II, the non-linear PB equation determining the electrostatic potential of this system, $\phi$, is~\cite{Frydel2007,Stillinger1961,Gray2018,Morozov2019} 
\begin{equation}\label{interfacePB}
\nabla\cdot \left(\epsilon(r,z)\nabla \tilde{\phi}\right) - \epsilon(r,z)\kappa^2(r,z) \sinh(\tilde{\phi}) = -\rho(r,z;z_c).
\end{equation}
Here, $\tilde{\phi}\equiv \frac{e_0\phi}{k_BT}$ is the dimensionless electrostatic potential, where $e_0$ is the elementary charge and $k_BT$ the thermal energy. We model the charge density of the virion, $\rho(r,z;z_c)$, as two oppositely-charged shells (see~Appendix for the precise functional forms used). For spherically symmetric virions, $\rho(r,z;z_c)=\rho(\sqrt{r^2+(z-z_c)^2})$ (Fig.~\ref{fig1}b). Finally, $\kappa(r,z)$ and $\epsilon(r,z)$ denote the spatially-varying inverse Debye length and dielectric permittivity respectively. 

The interfacial electrostatics depends on $\kappa_2/\kappa_1$ and the dielectric contrast $\epsilon_2/\epsilon_1$. (The properties of the virion interior, $\epsilon_3$ and $\kappa_3$, also affect the results, but they are not varied below.) The importance of non-linear effects is governed by the dimensionless charge density $\sigma^* = \frac{\sigma R e_0}{\epsilon_1 k_B T}$, which compares typical electrostatic  and thermal energies~\cite{Frydel2007}. Additional geometrical parameters are $\kappa_1 R$ and $\delta/R$. For an RNA virus with $R\sim 20-50$ nm, and a charge density $\sim 0.1-0.5$ $e_0/$nm$^2$ in the two shells, $\kappa_1 R \sim 20-50$, and $\sigma^*\sim 10-100$, whereas $\delta/R$ is $\sim 0.1$ is reasonable given molecular sizes of proteins and RNA. We vary $\kappa_2/\kappa_1$ and $\epsilon_2/\epsilon_1$ to model specific interfaces.

The electrostatic self free energy of the virion is obtained by summing $\tfrac{1}{2}\rho\phi$ over all space. In our cylindrical coordinates, Fig.~\ref{fig1}, it  depends only on $z_c$: 
\begin{equation}\label{selfenergy}
{\cal F}_{\rm elec}(z_c) = \pi \int_0^{+\infty} dr \int_{-\infty}^{+\infty} dz \, r \rho(r,z;z_c)\phi(r,z;z_c),
\end{equation} 
which we call the virion-interface `approach curve'. The electrostatic force exerted by the interface on the virion is given by $f(z_c)= - \frac{\partial {\cal F}_{\rm elec}(z_c)}{\partial z_c}$.

\begin{figure}[t]
\centerline{\includegraphics[width=0.5\textwidth, angle=0]{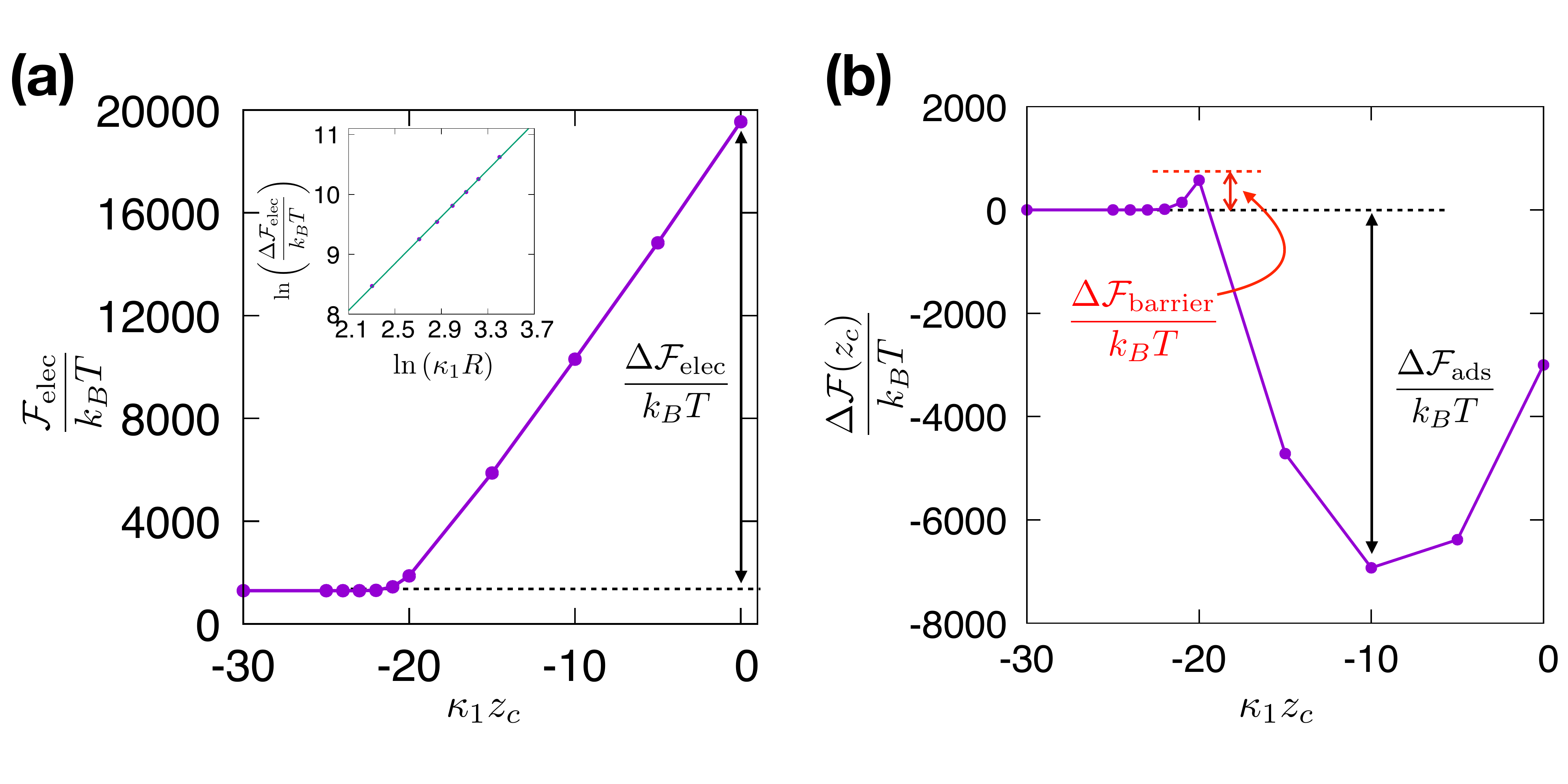}}
\caption{(a) Approach curve showing the increase in electrostatic free energy ${\cal F}_{\rm elec}(z_c)/k_BT$ as a function of distance to the air-water interface. Parameters are as specified in the text. Inset: scaling of the electrostatic free energy increase $\Delta {\cal F}_{\rm elec}={\cal F}_{\rm elec}(0)-{\cal F}_{\rm elec}(\kappa_1 z_c\to-\infty)$ in units of $k_B T$ as a function of dimensionless radius of the virion, $\kappa_1 R$. The line is a fit, and has slope $\sim 1.96$, which is close to the predicted value of $2$, see Appendix. (b) Plot of the total (electrostatic plus Pickering) free energy change, $\Delta {\cal F} (\kappa_1 z_c)={\cal F}(\kappa_1 z_c)-{\cal F}(\kappa_1 z_c\to-\infty)$, as a virion approaches an air-water interface ($\gamma=70$ mN/m). The electrostatic free energy barrier $\Delta {\cal F}_{\rm barrier}$ and the adsorption free energy gain $\Delta {\cal F}_{\rm ads}$ are shown in the plot.}
\label{fig2}
\end{figure}

\subsection*{Electrostatics provide a general physical mechanism for viral inactivation at air-water interfaces}

We first compute the self free energy of a viral particle approaching an air-water interface ($\kappa_2=0$, $\epsilon_2/\epsilon_1=1/80$). Typically we take $\kappa_1 R=20$, $\delta/R=0.125$ and $\sigma^*\simeq 17.2$ to be physiologically relevant. Approaching the interface increases the self free energy due to a generic build-up of electrostatic repulsion arising from the proximity of the air phase where there is no screening, and the permittivity is much smaller (Fig.~\ref{fig2}a). This increase is substantial, and is of the order of $10^4 k_BT$ for a viral particle close to the interface. The electrostatic free energy increase $\Delta {\cal F}_{\rm elec}={\cal F}_{\rm elec}(0)-{\cal F}_{\rm elec}(z_c\to-\infty)$ opposing adsorption scales as $R^2$ (Fig.~\ref{fig2}a, inset), as predicted by an analytic calculation (see below and Appendix for details). Remarkably, this scaling depends on the double-shell distribution typical of an RNA virus. For a single charged shell, representing an empty capsid, $\Delta {\cal F}_{\rm elec}$ instead scales as $R^3$, whereas if the genetic material fills the space inside the capsid uniformly, such as in a DNA virus like a bacteriophage, then $\Delta {\cal F}_{\rm elec}$ should scale as $R^5$.  

In the bulk, attractive van der Waals and hydrophobic interactions contribute a negative term of magnitude $\sim 10^4 k_BT$~\cite{Siber2012}, which balances out the electrostatic repulsive self energy and stabilises virions. The calculated free-energy increase on approaching an air-water interface (Fig.~\ref{fig2}a) is comparable in magnitude to attractive interactions. The total energy of a virus lodged at an interface may therefore become positive and trigger disassembly. Electrostatics is therefore  a generic physical mechanism for viral inactivation at air-water interfaces. 

The energy increase with decreasing $z_c$, so that viruses are strongly repelled electrostatically from an air-water interface on approach. However, when $|z_c| < R$, the virus breaches the interface and an additional component of the free energy 
must be considered. The virion now covers a circular part of the interface of area $\pi (R^2-z_c^2)$, reducing the total free energy. The corresponding `Pickering free energy'~\cite{Pickering1907,Ramsden1903} -- for a spherical viral particle breaching an interface with surface tension $\gamma$ -- can be estimated as
\begin{eqnarray}\label{FPick}
{\cal F}_{\rm Pick}(z_c) & = & -\pi \gamma (R^2-z_c^2), \qquad |z_c| \le R \\ \nonumber 
{\cal F}_{\rm Pick}(z_c) & = & 0, \qquad \qquad \qquad \, \, \, \, \, \, \, \, \, \, |z_c| \ge R.
\end{eqnarray}
For an air-water interface, $\gamma\sim 70$~mN/m, so that the Pickering free energy gain for the smallest RNA viruses with $R\sim20$~nm is already of the same order of magnitude as the electrostatic free energy increase computed in Fig.~\ref{fig2}a. Whether the Pickering or the electrostatic contribution wins then depends on parameter details, such as the exact charge density of the virion. For the case considered in Fig.~\ref{fig2}a, the minimum of the total free energy ${\cal F}={\cal F}_{\rm elec}+{\cal F}_{\rm Pick}$ occurs at $\kappa_1 z_c\simeq -10$ (Fig.~\ref{fig2}b), corresponding to an apparent contact angle (Fig.~\ref{fig1}a) of $\theta \sim \pi/3$ (or to $z_c/R\simeq -0.5$). The nontrivial $\theta$ ($\neq 0$ or $\pi/2$) is due to the different scaling of the electrostatic and Pickering free energies with $z_c$: the former is approximately linear (Fig.~\ref{fig2}a), the latter quadratic (Eq.~\ref{FPick}). 

Even for cases where the Pickering free energy gain is sufficient to favour adsorption (as in Fig.~\ref{fig2}), there is an energy barrier opposing this process. This barrier is purely electrostatic because it appears before the virion contacts the interface (when the Pickering contribution is zero), and is given by $\Delta {\cal F}_{\rm barrier}={\cal F}_{\rm elec}(-\kappa_1 R)-{\cal F}_{\rm elec}(\kappa_1 z_c \to -\infty)$, which is $\simeq 576$ $k_BT$ in Fig.~\ref{fig2}b. This barrier is too large for Brownian motion to overcome. Interestingly, experiments observe inactivation in viral suspensions following bubbling~\cite{Adams1948} or tumbling~\cite{Trouwborst1974}, suggesting that the process is not spontaneous, but indeed involves the non-thermal forces overcoming some free energy barrier. Inspection of  approach curves such as Fig.~\ref{fig2}a reveals that the force resisting adsorption and associated with the electrostatic free energy barrier is $\sim 0.1-1$~nN for typical viral parameters. To exert a viscous drag of this magnitude in a fluid of viscosity $\eta\sim 1$~cP, a velocity in the range of $v\sim 0.5-5$~ms$^{-1}$ is needed for virions with diameter $R=20$ nm, which is plausible in vigorous shaking or tumbling.


\subsection*{Interfacial and electrostatic forces determine the fate of a virion close to a liquid-liquid interface}

\begin{figure}[t]
\centerline{\includegraphics[width=0.5\textwidth, angle=0]{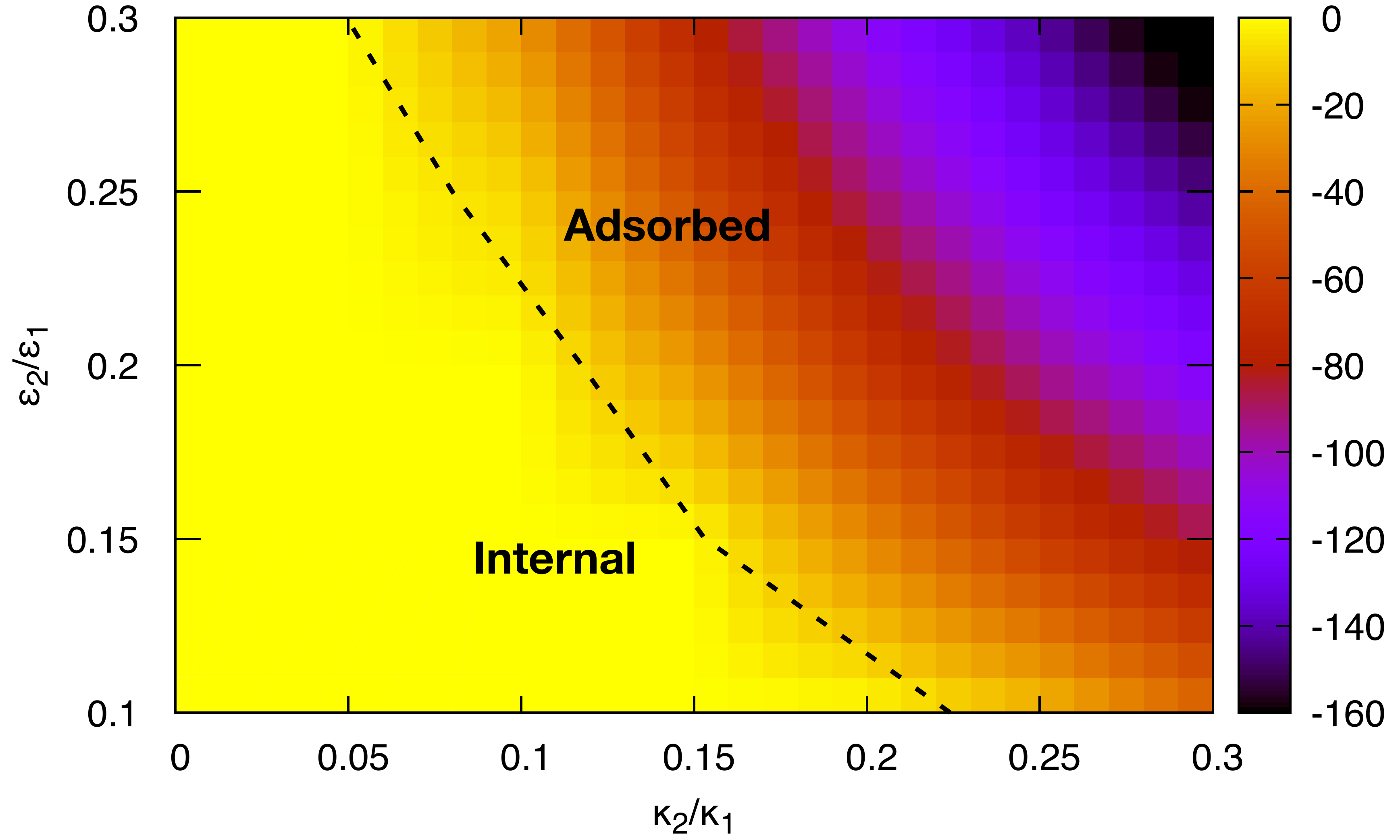}}
\caption{Phase diagram showing the fate of a viral particle approaching an interface between a physiological aqueous medium and another medium with variable electrostatic parameters. The heat map shows the adsorption free energy, ${\Delta \cal F}_{\rm ads}$, in units of $k_B T$, which includes both the electrostatic and Pickering contribution. The dashed line shows the location of the phase boundary between the phase where the system free energy is minimised with the virus in the aqueous phase (internal, $z_c<-R$) or at the interface (adsorbed, $z_c>-R$). Parameters are as in Figs.~\ref{fig1} and \ref{fig2}, except for $\gamma$, which is now $1.5$ mN/m to model the lower surface tension of liquid-liquid interfaces. }
\label{fig3}
\end{figure}

Consider now the balance between electrostatic cost and Pickering gain for different liquid-liquid interfaces (for which $\gamma$ is much lower, typically $\sim 1-10$ mN/m) as a function of $\kappa_2/\kappa_1$ and $\epsilon_2/\epsilon_1$. This question is of fundamental interest, as the case of $\kappa_2\ne 0$ was recently shown to be qualitatively different from that of $\kappa_2=0$ (relevant for air-water interfaces) as it leads to a distinct interparticle potential at the interface~\cite{Morozov2019,Muntz2020}. Figure~\ref{fig3} shows the total adsorption free energy ${\Delta \cal F}_{\rm ads}$ (defined in Fig.~\ref{fig2}b) for $0\le \kappa_2/\kappa_1 \le 0.3$ and $0.1\le \epsilon_2/\epsilon_1 \le 0.3$. The virus switches from being preferentially located in the aqueous phase at low $\kappa_2/\kappa_1$ or $\epsilon_2/\epsilon_1$ due to high electrostatic self energy, to being spontaneously adsorbed at large $\kappa_2/\kappa_1$ or $\epsilon_2/\epsilon_1$ due to the Pickering energy gain. 

To understand these results, we formulate a Debye-H\"uckel scaling theory valid for $\kappa_1 R \gg 1$ and $\delta/R \ll 1$, which is physically relevant for RNA viruses. We estimate ${\cal F}_{\rm elec}$ for a virion breaching the interface between fluids I and II as the sum of the self-energies of the two spherical caps in the two media (see Appendix), neglecting their interaction, which is numerically smaller because charges on the two halves are typically further apart than charges within each half. In this framework, the electrostatic self-energy of the virion at the interface is 
\begin{eqnarray}\label{theory}
{\cal F}_{\rm elec}(z_c) & \sim & \pi \sigma^2 R^2 \left(\frac{1-e^{-2\kappa_1 \delta}}{\epsilon_1 \kappa_1} + \frac{1-e^{-2\kappa_2 \delta}}{\epsilon_2 \kappa_2} \right) \\ \nonumber
  &+& \pi \sigma^2 R z_c \left(\frac{1-e^{-2\kappa_2 \delta}}{\epsilon_2 \kappa_2} - \frac{1-e^{-2\kappa_1 \delta}}{\epsilon_1 \kappa_1} \right).
\end{eqnarray}
Non-linear effects modify both the numerical coefficients and the dependency on $\kappa_1\delta$ or $\kappa_2\delta$ in Eq.~(\ref{theory}); neither does our simplified theory capture the presence of a non-zero electrostatic free energy barrier $\Delta {\cal F}_{\rm barrier}$ prior to interface contact. Thus, quantitative predictions require full PB numerics. Nevertheless, Eq.~(\ref{theory}) can explain qualitatively the results in Figs.~\ref{fig2} and \ref{fig3}. 

First, it predicts that ${\cal F}_{\rm elec}$ is linear in $z_c$ for all $\kappa_2$ and $\epsilon_2$, matching our numerics in Figs.~\ref{fig2} and S1. Secondly, Eqs.~(\ref{FPick}) and (\ref{theory}) predict (see Appendix) virion adsorption when 
\begin{equation}\label{phaseboundary}
\sigma^2 \left(\frac{1-e^{-2\kappa_2 \delta}}{\epsilon_2 \kappa_2} - \frac{1-e^{-2\kappa_1 \delta}}{\epsilon_1 \kappa_1} \right) -2\gamma \le 0,
\end{equation}
which gives a phase diagram in semi-quantitative agreement with Fig.~\ref{fig3} (see Fig.~S2). This equation also identifies the electrocapillary numbers $\frac{\sigma^2}{\gamma \epsilon_i \kappa_i}$ (with $i=1,2$), which measure the relative importance of electrostatic and interfacial effects in the two contacting media and determine the virion fate at an interface. 

To see the relevance of a transition between internal and adsorbed states (Fig.~\ref{fig3}) to disinfection, note first that respiratory RNA virions such as severe respiratory syndrome coronavirus 2 (SARS CoV-2) are borne by droplets rich in mucin, a very high molecular weight protein. As such droplets dry on surfaces, the mucin may form a gel state \cite{Marr2018} that, as in other hydrogels \cite{Klitzing}, remains permanently hydrated. When cleaning fluids contact such a composite object, there will be at least a transient interface between liquids of different compositions. As a test case, consider an interface between physiological saline and ethanol, the latter being a common ingredient in hand sanitizers.


In this case, $\kappa_2/\kappa_1\sim 0.05$, and $\epsilon_2/\epsilon_1\simeq 0.3$~\cite{Munday2008}, which is close to our predicted transition boundary between `internal' and `adsorbed' phases, while staying on the internal side for the parameters used for Fig.~\ref{fig3}. Importantly, the electrostatic $\Delta {\cal F}_{\rm barrier}$ is over an order of magnitude smaller than that associated with adsorption of the same virion onto an air-water interface (see Fig.~S1): i.e., the presence of ethanol renders it easier to drive virions close to an interface as the barrier is dramatically reduced. However, there remains enough dielectric contrast for inactivation through the increase in the electrostatic self-energy term once on the interface~\cite{note1}.
This may be one reason why alcohol is an efficient disinfectant. 


A related application is to a virion-laden droplet on skin. The relevant interfaces are now between physiological saline and sebum or sweat. The liquid components of sebum (triglycerides and squalene~\cite{Pappas}) have $\epsilon \gtrsim 2\epsilon_0$ and $\kappa \approx 0$ \cite{Benz}. We predict that viruses should stay inside the droplet and not be transferred into sebum. On the other hand, sweat is essentially a salt solution but at somewhat lower concentration than physiological saline~\cite{Keys}, so that virions should adsorb at the transient sweat-droplet interface. These predictions have evident implication for viral transmission via touching.  

We stress again here that the ability to fine tune interfacial parameters to determine the virus fate depends in large part on the fact that both Pickering and electrostatic contributions scale identically with the particle size, as $R^2$. This is only true for our double-shell charge distribution, typical of an RNA virus (Fig.~\ref{fig1}), and would not hold, for instance, for a bacteriophage where DNA fills the interior of the capsid approximately uniformly. Additionally, it is intriguing that van der Waals interactions, which keep the virion together, are also expected to scale as $R^2$~\cite{Siber2012}. It appears therefore that RNA viruses are uniquely placed at the edge of thermodynamic stability, which is reasonable for a system which needs to self assemble or disassemble on demand, in response to subtle changes in the surrounding medium. We suggest this very fact, though, makes it possible to design optimal disinfection strategies.




\subsection*{Discussion and conclusions}

In summary, we have computed the electrostatic free energy of an RNA virus approaching an interface between physiological saline and another medium. Our Poisson-Boltzmann formalism takes into account the spatial charge distribution of the virion and non-linear effects due to the highly charged nature of the virion's contituents. Our main finding is that at an air-water interface the electrostatic energy of a virus increases dramatically, by many thousands of $k_BT$, due to the low permittivity of and the absence of electrostatic screening in air. This provides an appealing physical mechanism to explain longstanding observations of viral inactivation at air-water interfaces \cite{Adams1948,Trouwborst1974,Thompson1999}. 

Including the Pickering free energy gain, which arises when any nanoparticle covers part of the interface between two media, leads to a transition between a regime where the viral particle is repelled from the interface to one where it adsorbs to it. Focussing on liquid-liquid interfaces, we find that varying the dielectric permittivities and Debye lengths of the two contacting media `tunes' which regime a given system is in. Potentially, then, disinfectants could, by design, be positioned close to the transition to give a low free energy barrier to adsorption. Virions in a droplets in contact with such a cleaning fluid may then adsorb, and therefore deactivate, more easily than when the droplet is in contact with air alone, where external energy is needed to achieve the same end~\cite{Adams1948,Trouwborst1974,Thompson1999}.

Our predictions are susceptible to experimental testing. Revisiting the classic viral deactivation experiments, e.g., of bubbling air through viral solutions, but now scanning the $(\epsilon_2/\epsilon_1, \kappa_2/\kappa_1)$ parameter space under carefully controlled conditions and detecting viral adsorption at interfaces would be one way forward. It should also be possible to compare the adsorption behaviour of RNA viruses with their empty counterparts that lack the genome \cite{Li2012}. We predict a significant difference in their electrostatic behavior near interfaces because the charge distribution of the latter is a single charged shell.


We have focussed on RNA viruses; but similar considerations should apply to DNA viruses such as bacteriophages, which are also  inactivated at interfaces~\cite{Trouwborst1974}. As anticipated, there will be qualitative differences, as the DNA of bacteriophages is arranged in a space-filling spool rather than on a thin shell, so that its electrostatic self-energy scales as $R^5$ rather than as $R^2$. As bacteriophages do not self-assemble but use a motor to package their genome, the electrostatic energy increase at the interface may result in DNA ejection rather than capsid disassembly~\cite{Trouwborst1974,Carrasco2009}. 

{\it Acknowledgements:} We thank M. Chiang for a critical reading of the manuscript and for useful discussions.

\newpage

\onecolumngrid

\section*{Appendix}

\subsection*{Poisson-Boltzmann numerics}

\begin{figure*}[!h]
\centerline{\includegraphics[width=0.6\textwidth, angle=0]{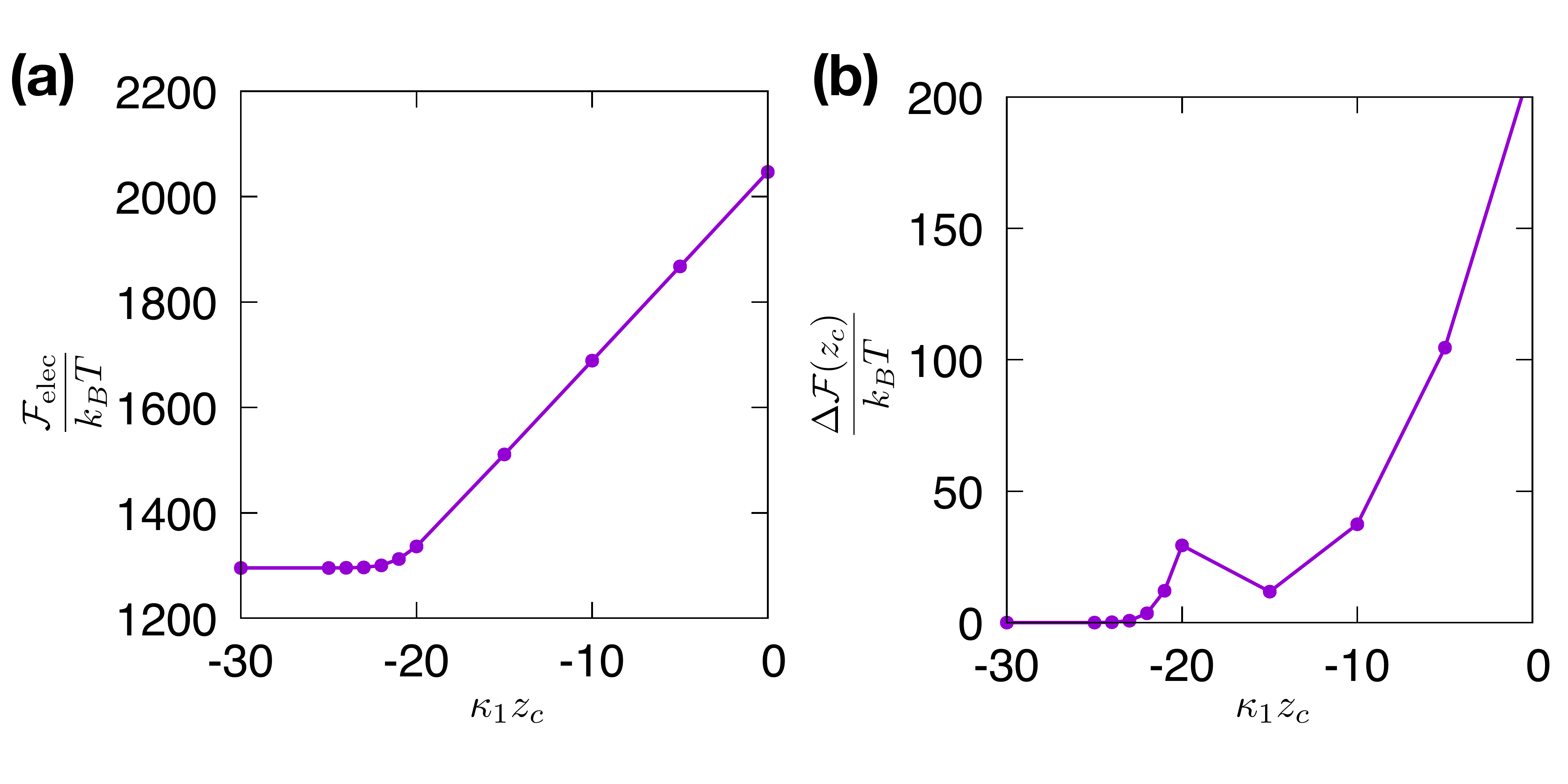}}
\caption{(a) Approach curve showing the increase in electrostatic free energy (or self-energy) ${\cal F}_{\rm elec}(z_c)$ as a function of distance to the interface, for two media with $\epsilon_2/\epsilon_1=0.3125$ and $\kappa_2/\kappa_1=0.05$. This is a typical situation for an interface between two media with different and finite Debye lengths (which is different to the case of an air-water interface where $\kappa_2=0$). (b) Plot of the total (electrostatic plus Pickering) free energy change, $\Delta {\cal F} (\kappa_1 z_c)={\cal F}(\kappa_1 z_c)-{\cal F}(\kappa_1 z_c\to-\infty)$, as a virion approaches the interface modelled in (a) (assuming a surface tension of $\gamma=1.5$ mN/m as in Fig.~3 of the main text). }
\label{figS1}
\end{figure*}

In the main text we show numerical solutions of the non-linear Poisson-Boltzmann (PB) equation for an RNA virion close to an interface between two media, characterised by different values of the dielectric constant and Debye length. The problem has cylindrical symmetry, and we call $z$ the position along the axis perpendicular to the interface ($z=0$ denotes the interface plane), and $r$ the distance to the centre of the viral particles in the plane parallel to the interface. In this geometry, the PB equation for monovalent salt electrolytes is,
\begin{equation}\label{interfacePBSI}
\nabla\cdot \left(\epsilon(r,z)\nabla \tilde{\phi}\right) - \epsilon(r,z)\kappa^2(r,z) \sinh(\tilde{\phi}) = -\rho(r,z)
\end{equation}
where $\tilde{\phi}\equiv \frac{e_0\phi}{k_BT}$ is the dimensionless electrostatic potential. To model an RNA virus, we considered a charge distribution $\rho(r,z)$ consisting of two concentric charged shells: an interior negatively charged shell (representing RNA) and an exterior positively charged one (representing the viral capsid, see Fig.~1 in the main text). The explicit functional form for $\rho(r,z)$ we considered is
\begin{eqnarray}
\rho(r,z) = \rho_0 \left[ e^{-\alpha (\tilde{r}-R-\delta)^2} - e^{-\alpha(\tilde{r}-R+\delta)^2}\right],
\end{eqnarray}
where $\tilde{r}=\sqrt{r^2+(z-z_c)^2}$, $R$ is the capsid radius, $2 \delta$ is the distance between the two charged shells, while  $\alpha$ controls the width of each shell. As mentioned in the main text, $\epsilon$ and $\kappa$ vary in medium I, II, and III (the capsid interior, see Fig.~1 in the main text); their values are respectively called $\epsilon_{1}$ and $\kappa_1$ (in medium I), $\epsilon_{2}$ and $\kappa_2$ (in medium II), $\epsilon_{3}$ and $\kappa_3$ (in medium III). In our numerical calculations we chose $\epsilon_1=80\epsilon_0$ (with $\epsilon_0$ the dielectric permittivity of vacuum) and $\kappa_1=1$ nm$^{-1}$ to model the aqueous phase containing the virion, $\epsilon_3=\epsilon_1/16=5\epsilon_0$ , and $\kappa_3=\kappa_1/10$ to model the virus interior, whereas we varied $\kappa_2$ and $\epsilon_2$. The case of an air-water interface corresponds to $\kappa_2=0$ and $\epsilon_2=\epsilon_0$ (Fig.~2 of the main text), whereas in Fig.~3 of the main text we varied $\kappa_2$ between $0$ and $0.3 \kappa_1$ and $\epsilon_2$ between $0.1 \epsilon_1$ and $0.3 \epsilon_1$ to model a liquid-liquid interface. The self-energy as a function of distance to the interface for a liquid-liquid interface with $\epsilon_2/\epsilon_1=0.3125$ and $\kappa_2/\kappa_1=0.05$ is shown in Fig.~\ref{figS1}, together with the total change in free energy (including the Pickering contribution): this case is relevant for the situation where medium II is an alcohol rub rich in ethanol.

In our numerical calculations we set $\kappa_1=\epsilon_1=1$, $R=20$, $\delta=2.5$, $\rho_0=0.3$, $\alpha=1$. Because $\phi$ is given in units of $k_BT/e_0$ as the potential in Eq.~(\ref{interfacePBSI}) has been made dimensionless, it follows that charge densities in Eq.~(\ref{interfacePBSI}) are measured in units of $k_B T \epsilon_1 \kappa_1^2/e_0$. With these choices, a free energy simulation unit equals $(k_BT)^2 \epsilon_1/(e_0^2 \kappa_1)$.

To solve Eq.~(\ref{interfacePBSI}), we used a finite difference scheme in cylindrical coordinates, with a $400\times 600$ grid, and a spatial discretisation of $\Delta x=0.5$ along both $r$ and $z$. We used a relaxation algorithm, introducing a time derivative so that Eq.~(\ref{interfacePBSI}) is solved in steady state. The time step in the relaxation algorithm was chosen to be $\Delta t=0.15\Delta x^2$, which is small enough for the algorithm to converge, yet large enough to not compromise computational efficiency.

\section{Self-energy calculations}

We now analyse the idealised case of systems made up by infinitesimally thin shells, and compute their self-energy in the Debye-H\"uckel approximation, which provides a useful baseline framework to interpret the results found by simulating the non-linear Poisson-Boltzmann equation. 

\subsection{Self-energies of single-shell systems}

It is useful to start with a single-shell system, which models an empty viral capsid. For a single shell of radius $R$, infinitesimal thickness, and surface charge density $\sigma$, the three-dimensional charge density entering the Poisson-Boltzmann equation can be simply written as
\begin{equation}
    \rho(r)=\sigma \delta(r-R),
\end{equation}
where $\delta$ here denotes the Dirac delta function.

Let us first analyse the case of $\kappa=0$ (no screening), which can be done by using standard electrostatics. We call $\epsilon$ the dielectric constant of the medium. In this case, the potential for $r\le R$ needs to be a constant, as the electric field is $0$ -- since there are no charges inside the shell. For $r>R$, the potential must be the same of that of a point charge with charge $Q=4\pi \sigma R^2$ (the total charge of the shell). Imposing continuity at $r=R$, we fix the constant and obtain that the electrostatic potential is given by
\begin{eqnarray}
\phi(r) & = & \frac{Q}{4\pi \epsilon R} \qquad r\le R \\ \nonumber
\phi(r) & = & \frac{Q}{4\pi \epsilon r} \qquad r > R.
\end{eqnarray}
As a consequence the self-energy for $\kappa=0$ is
\begin{equation}\label{selfenergy1}
{\cal F}_{\rm elec} = \frac{1}{2} Q\phi(R) = \frac{2\pi\sigma^2 R^3}{\epsilon}.
\end{equation}
Therefore for $\kappa=0$ the self-energy of a single shell scales as $R^3$.

Let us now turn to the case of a shell in a medium where there are mobile charges and electrostatic screening, $\kappa\ne 0$. One way to obtain the electrostatic potential in the Debye-H\"uckel approximation is to solve the linearised Poisson-Boltzmann equation for a charged shell,
\begin{equation}\label{DHsingleshell}
    \nabla^2 \phi -\kappa^2\phi = -\sigma \delta(r-R)/\epsilon.
\end{equation}
The general solution of Eq.~(\ref{DHsingleshell}) for $r\ne R$ is
\begin{equation}\label{generalsolutionDH}
    \phi(r)=A \frac{\cosh{(\kappa r)}}{r} + B \frac{\sinh{(\kappa r)}}{r}=
 \frac{A-B}{2} \frac{e^{-\kappa r}}{r} + \frac{A+B}{2} \frac{e^{\kappa r}}{r}
\end{equation}
For $r<R$, we need $A=0$ for the solution to be well behaved as $r\to 0$ (where there is no singularity, as there is no point charge, or $\delta$ function in the charge density). For $r>R$, we need $A+B=0$ as the potential needs to go to $0$ as $r\to\infty$. Integrating Eq.~(\ref{DHsingleshell}) over a thin shell around $r=R$ gives the discontinuity in $d\phi/dr$ (which is minus the electric field) at $r=R$ as
\begin{equation}\label{discontinuityDHsingleshell}
\left(\frac{d\phi}{dr}\right)_{r\to R^+}-\left(\frac{d\phi}{dr}\right)_{r\to R^-}=-\frac{Q}{4\pi \epsilon R^2}.
\end{equation}
This condition together with continuity at $r=R$ fixes the values of the constants so that the potential for $r\ge R$ is given by
\begin{equation}
    \phi(r)=\frac{Q \sinh{(\kappa R)}e^{-kr}}{4\pi\epsilon \kappa Rr},
\end{equation}
which may be viewed as the screened potential of a single point particle with charge $Q \sinh{(\kappa R)}/(\kappa R)$. The self-energy of the system is
\begin{equation}
    {\cal F}_{\rm elec} = \frac{1}{2} Q\phi(R) = \frac{2\pi\sigma^2 R^2}{\epsilon \kappa (1+{\rm cotanh}{(\kappa R)})},
\end{equation}
as found in~\cite{Siber2007,Siber2012}. For $\kappa R\gg 1$, which is relevant for viral capsid shells, the scaling therefore changes from $\sim R^3$ to $\sim R^2/\kappa$. 

\subsection{Self-energies of concentric shell systems}

In our cases the RNA virion is approximated by two concentric shells, of radius $R-\delta$ and $R+\delta$, and with charges $-\sigma$ and $\sigma$, representing the RNA and viral capsid respectively. Now the three-dimensional charge density entering the Poisson-Boltzmann equation is
\begin{equation}
    \rho(r)=\sigma \left[\delta(r-R-\delta)- \delta(r-R+\delta)\right].
\end{equation}
We call $Q_1=4\pi \sigma R_1^2$ and $Q_2=4\pi \sigma R_2^2$ the total charge of the inner and outer shell respectively.

Again, we consider the cases with $\kappa=0$ and $\kappa\ne 0$ separately, in analogy with the single-shell section. 

For $\kappa=0$, reasoning analogous to that in the previous section leads to the following ansatz for $\phi$,
\begin{eqnarray}
 \phi(r) & = & A \qquad \qquad \, \, \, \, \, \, \, r\le R_1\\ \nonumber
 \phi(r) & = & B+C/r \qquad R_1 < r \le R_2 \\ \nonumber
 \phi(r) & = & D/r \qquad \qquad r>R_2,
\end{eqnarray}
with $A$, $B$, $C$ and $D$ constants to be determined. Due to Gauss law (equivalently, by integrating the Poisson equation over spheres with radius just above $R_1$ and just above $R_2$), we find $C=\frac{Q_1}{4\pi\epsilon}$ and $D=\frac{Q_2-Q_1}{4\pi\epsilon}$. Requiring continuity at $r=R_1$ and $r=R_2$ fixes the remaining constants, yielding the following solution
\begin{eqnarray}
 \phi(r) & = & \frac{Q_2}{4\pi\epsilon R_2}-\frac{Q_1}{4\pi\epsilon R_1} \qquad r\le R_1 \\ \nonumber
 \phi(r) & = & -\frac{Q_1}{4\pi\epsilon r}+\frac{Q_2}{4\pi\epsilon R_2} \qquad R_1 < r \le R_2 \\ \nonumber
 \phi(r) & = & \frac{Q_2-Q_1}{4\pi\epsilon r} \qquad \qquad \, \, \, \, \, \, \, \, \, r>R_2.
\end{eqnarray}
The self-energy of two concentric shells with opposite surface charge density is therefore given by:
\begin{eqnarray}\label{selfenergytwoshellkappa0}
{\cal F}_{\rm elec} & = & \frac{1}{2}\left[-Q_1 \phi(R_1) + Q_2\phi(R_2) \right] \\ \nonumber
& = & \frac{Q_1}{8\pi\epsilon}\left(\frac{Q_1}{R_1}-\frac{Q_2}{R_2}\right)+\frac{Q_2 (Q_2-Q_1)}{8\pi \epsilon R_2} \\ \nonumber 
& = & \frac{4\pi\delta\sigma^2}{\epsilon}\left[-(R-\delta)^2+2R (R+\delta)\right].
\end{eqnarray}
From the last line of Eq.~(\ref{selfenergytwoshellkappa0}), it can be seen that, if $\delta\ll R$, then the self-energy of the two-shell system scales as $\sim R^2\delta$, which is different from the $\sim R^3$ scaling of the single-shell system found previously. The same calculations can be repeated for arbitrary charges in the two shells, and it is interesting that in the limit of $\delta \ll R$ the scaling remains the same if the charge densities in the two shells are tuned such that $Q_1=Q_2$ -- in this case the expression for the self-energy simplifies as the second term in the second line of Eq.~(\ref{selfenergytwoshellkappa0}) vanishes.

Let us now consider the case with screening, $\kappa\ne 0$. In each of the three regions $r< R_1$, $R_1<r<R_2$, $r>R_2$, the general solution is of the form given in Eq.~(\ref{generalsolutionDH}). As in the single shell calculation, the solution is proportional to $\frac{\sinh{(\kappa r)}}{r}$ for $r <R_1$ and to $\frac{e^{-\kappa r}}{r}$ for $r>R_2$, due to the boundary conditions at $r=0$ and $r\to\infty$. By integrating over infinitesimally small shells around $r=R_1$ and $r=R_2$ we obtain the conditions for the discontinuities of $\frac{d\phi}{dr}$,
\begin{eqnarray}
 \left(\frac{d\phi}{dr}\right)_{r\to R_1^+} - \left(\frac{d\phi}{dr}\right)_{r\to R_1^-}& = & \frac{Q_1}{4\pi \epsilon R_1^2} \\ \nonumber
 \left(\frac{d\phi}{dr}\right)_{r\to R_2^+} - \left(\frac{d\phi}{dr}\right)_{r \to R_2^{-}} & = & -\frac{Q_2}{4\pi \epsilon R_2^2}. 
\end{eqnarray}
Together with the continuity of $\phi$ at $r=R_1$ and $r=R_2$, these two conditions are sufficient to fix all remaining constants. The resulting potential for the two-shell system in the screened medium is
\begin{eqnarray}\label{potentialtwoshellkappane0}
 \phi(r) & = & \left(\frac{Q_2 e^{-\kappa R_2}}{4\pi\epsilon \kappa R_2}
-\frac{Q_1 e^{-\kappa R_1}}{4\pi\kappa\epsilon R_1}\right) \frac{\sinh{(\kappa r)}}{r} \qquad \qquad \qquad \qquad \qquad \qquad \qquad \qquad r\le R_1 \\ \nonumber
 \phi(r) & = & \left[\frac{Q_2 e^{-\kappa R_2}}{4\pi\epsilon \kappa R_2}+\frac{Q_1\sinh{(\kappa R_1)}}{4\pi \epsilon \kappa R_1}\right] \frac{\sinh{(\kappa r)}}{r} -\frac{Q_1 \sinh{(\kappa R_1)}}{4\pi\epsilon \kappa R_1} \frac{\cosh{(\kappa r)}}{r} \qquad \, \, \, \,  R_1 < r \le R_2 \\ \nonumber
 \phi(r) & = & \left[\frac{Q_2\sinh{(\kappa R_2)}}{4\pi\epsilon \kappa R_2}-\frac{Q_1 \sinh{(\kappa R_1)}}{4\pi\epsilon \kappa R_1}\right] \frac{e^{-\kappa r}}{r}\qquad \qquad \qquad \qquad \qquad \qquad \qquad \,   r>R_2.
\end{eqnarray}
It is interesting to note that the potential for $r>R_2$ is non-zero even if $Q_1=Q_2$ in a medium with screening (i.e.~$\kappa \neq 0$). In other words, the effective charge is a non-trivial combination of bare charges, screening length and geometric parameters. From Eq.~(\ref{potentialtwoshellkappane0}), the self-energy is found to be
\begin{equation}\label{selfenergytwoshellkappane0}
{\cal F}_{\rm elec} = \left[ \frac{Q_1^2 e^{-\kappa R_1}\sinh{(\kappa R_1)}}{8 \pi \epsilon \kappa R_1^2}+\frac{Q_2^2 e^{-\kappa R_2}\sinh{(\kappa R_2)}}{8\pi\epsilon \kappa R_2^2} - \frac{Q_1 Q_2 e^{-\kappa R_2}\sinh{(\kappa R_1)}}{4\pi\epsilon \kappa R_1 R_2}\right].
\end{equation}
In the limit $\kappa R \gg 1$, $\delta/R \ll 1$, which is the relevant one for RNA virions, we obtain
\begin{equation}\label{scaling1}
{\cal F}_{\rm elec} \sim \frac{2\pi \sigma^2 R^2}{\epsilon\kappa} \left(1-e^{-2\kappa \delta}\right).
\end{equation}
It should be noted that Eq.~(\ref{scaling1}) has a well defined limit for $\kappa\to 0$, which coincides, as it should, with the limit of Eq.~(\ref{selfenergytwoshellkappa0}) for $\delta/R\ll 1$.

\subsection{Scaling theory for an RNA virion at an interface}

We can start from Eq.~(\ref{scaling1}) to build a simple scaling theory for the free energy of an RNA virion at an interface between two media with different values of $\kappa$ and $\epsilon$, valid for $\kappa R\gg 1$ and $\delta/R \ll 1$. We begin by noting that Eq.~(\ref{scaling1}) can be written as a contribution proportional to the virion surface embedded in the medium, $S$, as
\begin{equation}\label{approximationsum}
    {\cal F}_{\rm elec} \sim \frac{S \sigma^2}{2 \epsilon \kappa} \left(1-e^{-2\kappa \delta}\right),
\end{equation}
which gives an expression for the electrostatic free energy per unit area, ${\cal F}_{\rm elec}/S$.  When the virion is at the interface, Fig.~1 in the main text, there are two spherical caps, one in each medium and with surface areas $S_1(z_c) = 2\pi R(R-z_c)$ and $S_2(z_c)= 4\pi R^2 - S_1(z_c)$ for $|z_c| \leq R$, with $z_c$ the interfacial height. We approximate the electrostatic free energy as the sum of the contribution of the two surfaces:
\begin{eqnarray}\label{scaling2}
  {\cal F}_{\rm elec} & \sim & \frac{S_1 \sigma^2}{2 \epsilon_1 \kappa_1} \left(1-e^{-2\kappa_1 \delta}\right) +\frac{S_2 \sigma^2}{2 \epsilon_2 \kappa_2} \left(1-e^{-2\kappa_2 \delta}\right) \\ \nonumber
  & = & \pi \sigma^2 R^2 \left(\frac{1-e^{-2\kappa_1 \delta}}{\epsilon_1 \kappa_1} + \frac{1-e^{-2\kappa_2 \delta}}{\epsilon_2 \kappa_2} \right) 
  + \pi \sigma^2 R z_c \left(\frac{1-e^{-2\kappa_2 \delta}}{\epsilon_2 \kappa_2} - \frac{1-e^{-2\kappa_1 \delta}}{\epsilon_1 \kappa_1} \right).
  \end{eqnarray}
If we estimate the total free energy of an RNA virion close to an interface as ${\cal F}(z_c)={\cal F}_{\rm elec}(z_c)+{\cal F}_{\rm Pick}(z_c)$, with ${\cal F}_{\rm Pick}$ the Pickering contribution discussed in the main text, and explicitly given by
\begin{eqnarray}\label{FPick}
{\cal F}_{\rm Pick}(z_c) & = & -\pi \gamma (R^2-z_c^2), \qquad |z_c| \le R \\ \nonumber 
{\cal F}_{\rm Pick}(z_c) & = & 0, \qquad \qquad \qquad \, \, \, \, \, \, \, \, \, \, |z_c| \ge R, 
\end{eqnarray}
 with $\gamma$ the surface tension between the two media, then the minimum of ${\cal F}$ is obtained with $z_c=z_c^*$, with
 \begin{equation}\label{scaling3}
 z_c^*=-\frac{\sigma^2 R}{2\gamma}  \left(\frac{1-e^{-2\kappa_2 \delta}}{\epsilon_2 \kappa_2} - \frac{1-e^{-2\kappa_1 \delta}}{\epsilon_1 \kappa_1} \right).
 \end{equation}
Consequently, the adsorption free energy is
  \begin{eqnarray}\label{adsorptionfreeenergy}
 {\Delta \cal F} & = & {\cal F}(z_c=z_c^*)-{\cal F}(z_c\to -\infty) \\ \nonumber
 & = & \pi R^2 \left[\sigma^2 \left(\frac{1-e^{-2\kappa_2 \delta}}{\epsilon_2 \kappa_2} - \frac{1-e^{-2\kappa_1 \delta}}{\epsilon_1 \kappa_1} \right) -\gamma -\frac{\sigma^4}{4\gamma} \left(\frac{1-e^{-2\kappa_2 \delta}}{\epsilon_2 \kappa_2} - \frac{1-e^{-2\kappa_1 \delta}}{\epsilon_1 \kappa_1} \right)^2\right],
 \end{eqnarray}

\begin{figure}[!h]
\centerline{\includegraphics[width=0.5\textwidth, angle=0]{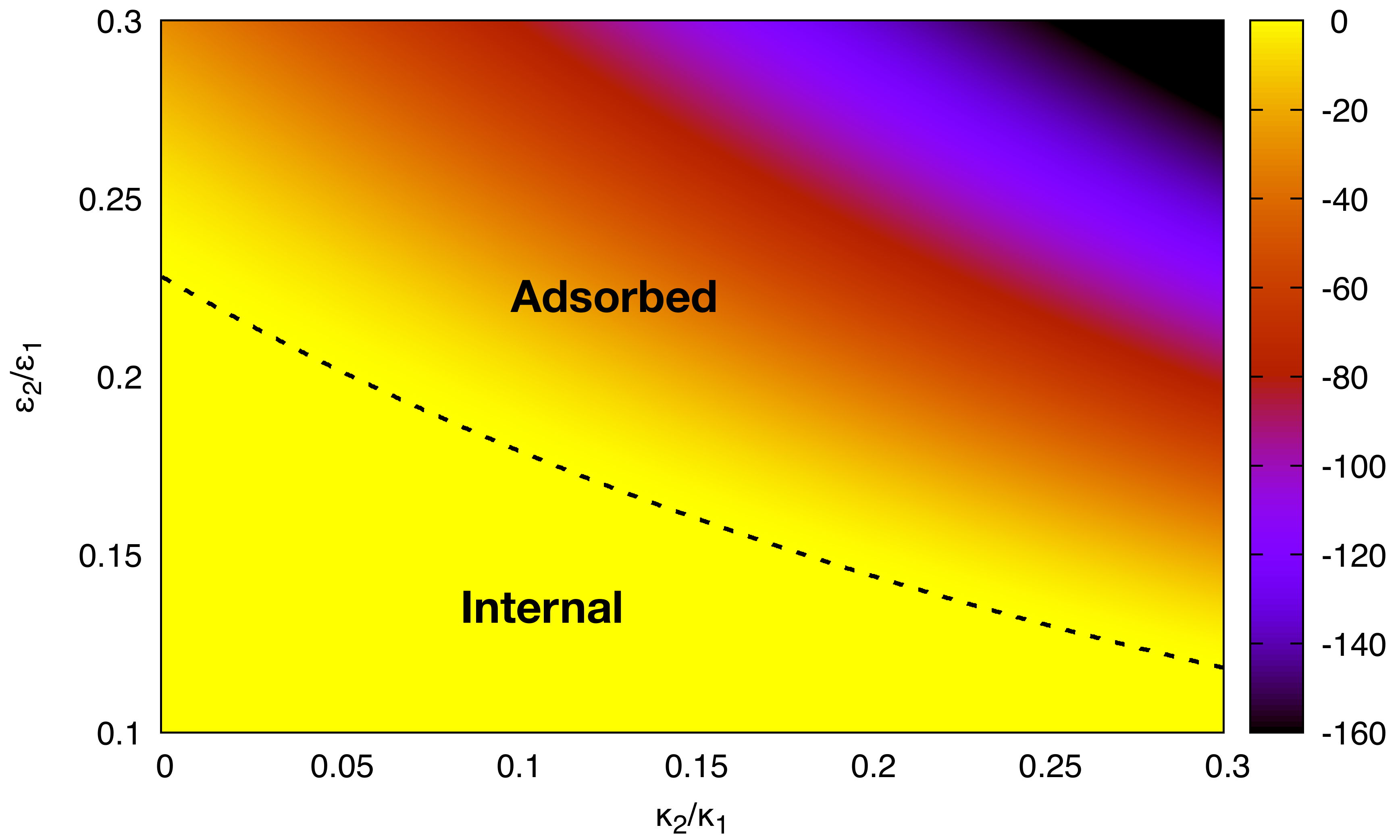}}
\caption{Phase diagram showing the fate of a viral particle approaching an interface between a physiological aqueous medium and another medium with variable electrostatic parameters, found by using our scaling theory. The heatmap shows the value of the adsorption free energy, Eq.~(\ref{adsorptionfreeenergy}), in units of $k_BT$. The dashed line shows the line where the adsorption free energy is zero (equivalently $z_c^*=-R$). The phase diagram in the Figure corresponds to $\sigma^*=11$, $\kappa_1=1$ nm$^{-1}$, $\delta=2.5$ nm, $\epsilon_1=80 \epsilon_0$, $\gamma=1.5$ mN/m.}
\label{figS2}
\end{figure}

Figure~\ref{figS2} shows the phase diagram found for a viral particle close to an interface between two media, found by our scaling theory (see caption for parameter list). The region above the dashed line in Fig.~\ref{figS2} corresponds to the adsorbed phase, the one below to the internal phase. The phase diagram is semi-quantitatively similar to the one shown in the main text found by PB numerics, which, unlike this approximate treatment, capture non-linear electrostatics. 

Our scaling theory can be generalised for charge distributions other than the two-shell system modelling an RNA virus. For a single charged shell, modelling an empty viral capsid, at the air-water interface, proceeding as in Eq.~(\ref{approximationsum}), we can equate the self-energy of the system to the sum of the self-energies of two spherical caps. The self-energy of the cap in water scales as its surface, whereas the self-energy of the cap in air scales as the volume [see Eq.~(\ref{selfenergy1})]. As a result, we expect the adsorption free energy for an empty capsid should scale as $R^3$, as the contribution of the sphere cap in air dominates for large $R$. This means that the Pickering free energy gain cannot offset the electrostatic cost of reaching the interface as $R$ increases. 

For a uniformly charged sphere, modelling the DNA spool inside a bacteriophage, the self-energy scales as $R^5$ in air (or a medium with $\kappa=0$) and as $R^3/\kappa^2$ in a medium with $\kappa\ne 0$~\cite{Siber2007,Siber2012}. Consequently, we expect the adsorption free energy to scale as $R^5$ for an air-water interface and as $R^3$ for a liquid-liquid interface. In both cases, the predicted increase with $R$ of the adsorption free energy is steeper than for two-shell RNA virions. 


\begin{thebibliography}{99}
\bibitem{note1}{Indeed, a disassembled virion may be particularly stable at the interface, because the Pickering free energy gain remains, while the large electrostatic free energy cost from charge confinement inside the capsid vanishes.}.
\end{thebibliography}
\end{document}